\documentclass[10pt,conference]{IEEEtran}
\IEEEoverridecommandlockouts
\usepackage{cite}
\usepackage{amsmath,amssymb,amsfonts}
\usepackage{algorithmic}
\usepackage{graphicx}
\usepackage{textcomp}
\usepackage{xcolor}
\usepackage{float}
\usepackage{adjustbox}
\usepackage{xspace}

\usepackage{booktabs}
\usepackage[normalem]{ulem} 
\usepackage[autostyle]{csquotes}

\usepackage{ifthen}
\usepackage{amssymb}
\usepackage{adjustbox}

\newcommand\devops{DevOps}
\newcommand\numberofrepositories{148\xspace}
\newcommand\numberofprojects{72\xspace} 

\newcommand\totalnumberofstudents{352\xspace} 
\newcommand\totalnumberofstudentsdevops{240\xspace}
\newcommand\totalnumberofquestionnaire{86\xspace}
\newcommand\coursetime{four\xspace}

\usepackage{hyperref}
\hypersetup{colorlinks,allcolors=blue}

\usepackage{breakurl}

\newcommand\organizacagit{https://github.com/fga-eps-mds}

\def\BibTeX{{\rm B\kern-.05em{\sc i\kern-.025em b}\kern-.08em
    T\kern-.1667em\lower.7ex\hbox{E}\kern-.125emX}}
    
\begin{document}

\title{Qualifying Software Engineers Undergraduates in DevOps - Challenges of Introducing Technical and Non-technical Concepts in a Project-oriented Course\\
}

\author{\IEEEauthorblockN{Isaque Alves}
\IEEEauthorblockA{
\textit{University of Brasília (UnB)}\\
Brasília, Brasil \\
isaquealvesdl@gmail.com}
\and
\IEEEauthorblockN{Carla Rocha}
\IEEEauthorblockA{
\textit{University of Brasília (UnB)}\\
Brasília, Brasil \\
caguiar@unb.br}
}

\maketitle

\begin{abstract}
The constant changes in the software industry, practices, and methodologies impose challenges to teaching and learning current software engineering concepts and skills. \devops\xspace is particularly challenging because it covers technical concepts, such as pipeline automation, and non-technical ones, such as team roles and project management. 
The present study investigates a course setup to introduce these concepts to software engineering undergraduates. We designed the course by employing coding to associate \devops\xspace concepts to Agile, Lean, and Open source practices and tools. We present the main aspects of this project-oriented \devops\xspace course, with  \totalnumberofstudentsdevops students enrolled it since its first offering in 2016. 
We conducted an empirical study, with both a quantitative and qualitative analysis, to evaluate this project-oriented course setup. We collected the data from the projects repository and students' perceptions from a questionnaire. We mined \numberofrepositories repositories (corresponding to \numberofprojects projects) and obtained \totalnumberofquestionnaire valid responses to the questionnaire. We also mapped the concepts which are more challenging to students learn from experience.
The results evidence that first-hand experience facilitates the comprehension of \devops\xspace concepts and enriches classes discussions.  we present a set of lessons learned, which may help professors better design and conduct project-oriented courses to cover DevOps concepts. 
\end{abstract}



\section{Introduction}
\label{SecIntroduction}

The development rate of the software industry has evolved from \textit{delivering} the software product in the waterfall model to \textit{continuous delivering} in agile and \devops\xspace \cite{article:kent:2013}. This constant evolution of software industry standards, practices, and methodologies imposes challenges to teaching and learning software engineering \cite{hopper2012practicing}\cite{pang2020}. Education programs must connect abstract concepts taught in the classroom to skills needed for software engineering practitioners \cite{stackoverflow2017earn}\cite{beyer2016site}\cite{microsoft2018devops} \cite{broman2012company} \cite{vanhanen2012teaching}. However, it is common to see  undergraduate software engineering courses  covering contents and concepts disconnected from current industry practices.  

In this context, teaching \devops\xspace is exceptionally challenging. \devops\xspace  is a natural evolution of the agile movement \cite{Christensen:2016}\cite{humble2010continuous}, and it proposes a complementary set of  practices and automation to allow the iterative delivery of software in short cycles effectively \cite{Leo2019}\cite{travassos2016SLR}\cite{rong2017devopsenvy}.  To understand the factors that impact frequent and reliable release process, the student should understand concepts such as continuous improvement, product management over project management, small features, measurement, change, risk cost, short feedback cycle, communication, a culture of collaboration, and cross-functional/full-stack team, among others. Additionally, mastering technical concepts are essential to enable a continuous deployment pipeline.  Deployment pipeline, continuous deployment, versioning, containerization, build automation, static code analysis, continuous integration, testing automation, microservice architecture, automation tools, cloud services, and continuous runtime monitoring guarantee supports a stable and repeatable deployment process. However, it is far from trivial to bring to a classroom context the complexity and the interdependence between these concepts and their practical implications. How to qualify engineers for \devops\xspace practice is still an open research question \cite{Leo2019}\cite{pang2020}.

A commonly adopted strategy in software engineering courses to present complex concepts and develop technical students' skills is to foster them to participate in Open Source Software (OSS) projects as part of the course \cite{Holmes:2018}\cite{hu2018}\cite{Pinto:2019}. By working with OSS projects, students learn skills, practices, toolsets, and technologies aligned with the current software development landscape \cite{Nascimento2013UsingOS}. Such a scenario also enhances students' resumes and parallels software engineering concepts with real-world scenarios.

This paper presents a  project-oriented course setup introducing both technical and non-technical \devops\xspace concepts to undergraduates.  We employ coding to associate \devops\xspace concepts to Agile, Lean, and OSS practices and tools. Throughout the course, teams contribute to a capstone OSS project, adopting these proposed practices in the project development cycle.   Many OSS communities have embraced DevOps tools and practices. We adopt many of these automation tools in the projects. In the classroom, the professor introduces \devops\xspace theoretical concepts, correlates them with these practices, and presents how they impact continuous delivery.  

The insights obtained from this empirical research provide lessons learned for professors, students, and OSS communities that want to take advantage of this process. The main contributions of this papers are the following:

\begin{itemize}
    \item We present a project-oriented course setup that associates \devops\xspace concepts with practices and tools from Open Source Community, Agile and Lean. Previous work on the topic focused mostly on tools and automation of the deployment pipeline, and \devops\xspace still imposes challenges due to its complexity \cite{Leo2019}. We could not find a research paper that covers the teaching challenges of presenting the complexity of \devops\xspace \cite{pang2020}, to give an understanding that non-technical aspects, such as communication and management, together with technical aspects, such as pipeline and provision automation, impacts the continuous software delivery;
 
   \item From the course data analysis, we present a set of recommendations/lessons learned to guide professors, instructors, and practitioners. Based on the course data analysis, we present a set of lessons learned, which may help professors better design and conduct project-oriented courses to cover DevOps concepts. 
\end{itemize}   

The rest of the paper is organized as follows. In Section \ref{SecMethod} we present our research question and the research methodology. Section \ref{SecResearchDesign} presents the course setup. In Section \ref{QuantitativeAnalysis}, we present a quantitative analysis with data mining from the course repository. Finally, in Section \ref{SecFurtherAnalysis}, we present our lessons learned from \coursetime years of course experience, and the related works in Section \ref{SecRelatedWorks}. The conclusions are in Section \ref{SecConclusion}.

\section{Method}
\label{SecMethod}

In this section, we state the research question and the process followed in conducting data analysis. We defined one research question that guided our study:

\textbf{RQ.} \textit{Can a project-oriented course help to teach both technical and non-technical \devops\xspace concepts to undergraduates?}

This research question guides us to plan, execute, and evaluate a project-oriented course focused on \devops. We conducted empirical research to answer this research question. First, we assigned practices and tools from agile, lean, and OSS to \devops\xspace concepts. It guides us to plan the presented project-oriented course.

To evaluate the course syllabus and the relevance of these practices, we mined data over \numberofrepositories repositories from \numberofprojects projects extracted from the Github organization created for the course, from all teams enrolled in the course. We take data over \coursetime years in order to compare the \devops\xspace and agile course strategies.  In this period, more than \totalnumberofstudents undergraduate software engineering students attended it. Finally, we conducted a questionnaire to evaluate the students' perceptions.

\subsection{Assigning practices to concepts}
\label{Sub:assignPracticestoConcepts}

The course, called ``Software Product Development'' (SPD), is offered to undergraduate software engineers, twice a year (Fall and Spring). The course was first offered in 2012, adapted from the XP lab \cite{goldman2004}, similar to \cite{jari2012}, and focusing on agile practices. In 2014, we introduced OSS practices, and in 2016, the course was reformulated to focus on \devops. 

To create the course syllabus, we adopted the \devops\xspace conceptual map from the survey \cite{Leo2019},  organized into four categories: process, people, delivery, and runtime. Each category presents a set of concepts and how they relate to each other in the \devops\xspace perspective. \textit{Delivery} and \textit{Runtime} categories cover mostly technical aspects of \devops, including: Continuous delivery/deployment, deployment pipeline stages, automation, open source, versioning, build management, static analysis, testing automation, continuous integration, build once, configuration management, microservices, rollback, continuous runtime monitoring, infrastructure as code, containerization, cloud services, environments are similar, performance, availability, scalability, resilience, reliability, less human intervention. 

\textit{People} and \textit{Process} categories cover concepts for project management and collaboration, including: frequent and reliable release process, agile/lean, short feedback cycle, product quality, small features, innovation, trust relationships, continuous improvements, product management over project management, risk and cost, a culture of collaboration, \devops\xspace role, aligning incentives, personal responsibility, failure as an opportunity for improvement, cross-functional, "you build it, you run it", deployment triggered by dev, knowledge, tools, processes, and practices. 

We excluded the concepts related to software maintenance and evolution, like feature toggle, run time experiments, A/B testing, Canary release, chaos engineering, log management, and continuous post-production maintenance. They require advanced projects, and it is out of the scope of a capstone project lifecycle. From this first filter, we obtained the list of \devops\xspace concepts related to the software project development cycle, as depicted in Table \ref{tab:commands}.

Then, we listed agile, OSS, lean, and \devops\xspace practices; we also listed the open source tools adopted by \devops\xspace practitioners community to encourage students to use tools adopted in the industry. We employed coding to assign each practice to at least one \devops\xspace concept \cite{stol2016grounded}.  From codes to categories\cite{charmaz2008grounded}: based on the initial list of practices (the codes), we analyzed to identify the category of these practices. In the following, we present some examples of concepts and related practices:

\begin{itemize}

\item \textit{Culture of collaboration} - core to \devops, is based on sharing knowledge, tools, processes and practices \cite{Leo2019}. Practices that promote this concept, when done with a full-stack, self-organized team, and a centralized repository: stand-up meeting, communication in issues and pull requests,  pair revision, and planning; 

\item \textit{Continuous delivery/deployment} - we employ deployment automation tools to make the continuous delivery process viable \cite{Leo2019}\cite{keefe2004using}\cite{rong2017devopsenvy}. Practices that enables continuous deployment: configure automated deployment pipelines/stages, unit tests, integration tests, continuous integration, automate build and deploy process;

\end{itemize}

\subsection{Repository Analysis}

We extracted the data from the GitHub repositories at the course organization\footnote{\url{\organizacagit}}. The goal is to understand how students employ \devops\xspace concepts in the project. This organization contains the projects from 2014 to the present, and more than \totalnumberofstudents students have enrolled in this course. For this study, we considered all \numberofprojects projects. 

We extracted several metrics/data from each repository: the number of issues open/closed per sprint, the number of pull requests open/closed per sprint, the number of stages per pipeline, tools used, technical/project documentation, and languages/frameworks used. They are available, and some of these quantitative data reflect non-technical practices. For instance, team communication is directly related to the frequency they interact in the issues, the number of closed issues, and interactions in pull requests.
The qualitative data in the repository, such as comments on issues and project documentation, are considered in students' evaluation. However, we did not use this data in the present study due to its subjectivity. 

\subsection{Questionnaire}

To understand the students' perspective on the course methodology and the \devops\xspace concepts, we designed an online questionnaire and submitted it to students that took this course from August 2016 to June 2019. The goal of this questionnaire is to verify if students could perceive the complexities and correlate technical with non-technical aspects of  \devops.  The survey consisted of 65 Likert scale questions on their perception of the \devops\xspace concepts, the agile, lean and OSS practices, and open-ended questions eliciting the respondent's opinion in general \cite{Punter2003}. 

We received \totalnumberofquestionnaire valid responses to our survey. We analyzed the data using descriptive statistics and open coding of textual comments. To reduce the subjectivity of this process,  both authors analyzed the data. We also provide representative quotes to highlight our findings. 

\section{The Course Setup}
\label{SecResearchDesign}

\begin{table*}[hbt]
\caption{\devops\xspace Concepts and practices presented in the course.}
 \label{tab:commands}
\resizebox{\textwidth}{!}{%
\begin{tabular}{p{.15\textwidth}|p{.3\textwidth}|p{.3\textwidth}|p{.23\textwidth}}
\hline
\textbf{Categories}   & \textbf{Concept} & \textbf{Practice} & \textbf{Tools} \\ 
\hline
Product Management 
\newline (Process/ People) 
& Microservice; Product quality;\newline Customer satisfaction; Small features; \newline Artifact management; Release engineering; \newline Knowledge, Skills, and Capabilities;  \newline Programing educations; Quality assurance; \newline and Artifact management.
& OSS Documentation standards; \newline unit test; review; licensing; \newline pair revision and code review; \newline architecture structure; \newline product documentation; and \newline pipeline stage documentation. 
& wiki; git Pages; codeclimate; \newline coveralls; git; and SonarQube.
\\
\hline

Project Management \newline (Process/ People) 
& ZenHub; HubCare; wiki; \newline People management; Short feedback cycle; \newline Pilots team and lead customer; \newline Compliance regulations; Team experience; \newline Aligning incentives; Breaking down silos; \newline Culture of collaboration; Versioning; \newline Sharing knowledge; Programing educations; \newline Global community knowledge; \newline Failure as opportunity of improvement; \newline ollaborate across departments; \newline Knowledge, skills, and capabilities; and \newline Artifact management.
& Sprint; Kanban; planing; review; \newline Stand-up metting; dojos; \newline Tasks in issues; training; \newline OSS Recommended Standards; \newline Post mortem; code of contributing; \newline Communication in issue and PR; \newline Full-stack and self organized teams; \newline git-flow; pull request; process \newline documentation; badges status \newline in readme; tracking metrics; and \newline pair revision. 
& ZenHub; HubCare; wiki; \newline Agile Visualization; Telegram; \newline Slack; Rocket Chat; Redmine; \newline Google Drive; programming \newline language; and test tools.
\\ 
\hline

Build Process \newline (Delivery) 
& Release engineering; Continuous delivery; \newline Automation; Testing automation; \newline Correctness; and Static analysis.
& Deployment pipeline stages; \newline Automation; Unit test; Integration \newline tests; and components tests.
& AWS Cloud Formation; Jest; \newline Jmeter; Selenium; Sonar; Rancher; \newline Cucumber; and Code Climate.
\\ 
\hline

Continuous Integration \newline (Delivery) 
& Frequent and reliable release process; \newline Release engineering; Deployment pipeline; \newline Continuous integration; Automation; \newline and Continuous delivery.
& Build and deploy automated; \newline badges status in readme; \newline and git-flow.
& Jenkins; Gitlab CI; \newline Circle CI; and Travis.
\\ 
\hline

Deployment Automation \newline (Delivery/Runtime) 
& Frequent and reliable release process; \newline Release engineering; Continuous delivery; \newline Configuration management; Automation; \newline Infrastructure as code; Virtualization; \newline and Containerization.
& Git-flow; Continuous Integration; \newline Build and Deploy automated; \newline Architecture Structure; \newline Documentation; and pipline stage \newline documentation.
& Shell Script; Travis; Docker; npm; \newline Virtualenv; OpenStack; Vagrant; \newline Kubernets; Cloud services; Chef; \newline Puppet; Heroku; Flyway; Rancher; \newline and AWS Cloud Formation.
\\ 
\hline

Monitoring \& Logging \newline (Delivery/Runtime) 
& You built it, you run it; Availability; \newline After-hours support for Devs; Alerting; \newline Continuous runtime monitoring; Security; \newline Performance; Automation Metrics; \newline Experiments Log management; Reliability; \newline and Scalability Resilience.
& Operation tasks in issues; \newline Logging; Monitoring; and \newline Benchmark.
& Prometheus; Graylog; Arachni; \newline Nagios; Zabbix; Logstash; \newline Graylog; and ElasticSearch.
\\ 
\hline

\end{tabular}
}
\end{table*}
\subsection{Course Overview}

The course has over 17 weeks (136 hours), with a theoretical and hands-on class twice a week. Additionally, it requires, on average, ten weekly hours of student efforts extra class, with students dedicating to their OSS project development. This course admits students with intermediate programming skills, with experience in requirement engineering, software architecture, code refactoring techniques, and basic knowledge of software project lifecycle. The complete set of concepts covered in this course, the corresponding practices, using the process described in \ref{Sub:assignPracticestoConcepts} are depicted in Table \ref{tab:commands}. This table guides professors, tutors, and students throughout the project lifecycle.

In the initial course weeks, students organize themselves in teams and apply to one of the available open problems. The professor is responsible for validating/assigning a theme to each team. In the classroom, the professor introduces the \devops\xspace concepts, and practices depicted in Table \ref{tab:commands}. In parallel, the teams execute a software project development cycle employing all practices from Table \ref{tab:commands}. 
In the planning stage, they learn more about the product they will develop, organize their repository, and perform training dojos with mentors' help. In the analysis stage, they start requirement elicitation, project architecture definition, pipeline tools definition, and roadmap of their deliveries. 

In the design stage, students propose the pipeline stages, automation tools, frameworks, to ensure continuous delivery. Risk management, project quality, and deliverables are planned , and students monitor metrics and indicators to track productivity. For seven weeks, they design and implement some features of their project. At the end of this period, we have a first release of their prototype, and  the professor gives feedback and assess risks. The remaining ten weeks are dedicated to the continuous delivery and deployment, with students deploying at least once a week a new version of their  Proof of  Concept, guaranteeing the manual usability tests, continuous deployment pipeline, unit test coverage, and other test stages. The final delivery contemplates the software delivery, the technical and methodology documentation, and the compliance with OSS recommended standards.

It is students' first experience leading a team, making decisions, planning releases, and implementing a continuous deployment pipeline.  Students will execute the practices from Table \ref{tab:commands} throughout the project lifecycle, make decisions on tools, manage a team composed of 10 members, plan release cycles, document project execution and product management, and justify their decision-making. 

\subsection{Full-stack Team}

One common understanding between practitioners, managers, and organizations is that the \devops\xspace movement promotes closer collaboration between all team members (developers, operators, managers) \cite{dyck2015release} \cite{feijter2018maturity}. This close collaboration consists of sharing of knowledge and tools, processes, and practices, risks, and the culture \textit{"You built it, you run it"} \cite{Leo2019}.

To introduce this core concept, the team must be accountable for executing practices in Table \ref{tab:commands}. Therefore, each student in the team assumes one specific role throughout the project. The assigned student must study the concepts, learn specific tools, and assume the responsibility to guarantee the execution of the practices and the deliverable related to the role. Each team member develops distinct skills, focusing on one or more categories from Table \ref{tab:commands}, pushing them to communicate effectively and collaborate to guarantee the project delivery. The roles, their attributions, and deliverable are:
\begin{itemize}
   \item \textbf{Scrum Master/Tech Leader}: responsible for executing the concepts and practices from the \textit{Project Management} category. The expected deliveries are the process documentation, productivity metrics, and continuous analysis, evaluation of teams agile practices maturity; 
    \item \textbf{Product Strategist}: also responsible for executing the concepts and practices from the \textit{Product Management} category. This role must describe the user personas, how the product fits into the current market, and how it will achieve business goals. The expected deliveries are product plans, roadmap, business plan, usability tests, and visual product identity. 
    \item \textbf{Software Architect}: responsible for executing the concepts and practices from \textit{Product Management} and \textit{Build Process} categories. It deals with the application and its data flow by defining structures, tools, and technologies. The expected deliveries are architecture definition and documentation, programming language selection, external services (Open Source) usage, functionalities reuse, and services integration, defining the quality criteria.
    \item \textbf{Release/devops\xspace Engineer}: responsible for the execution of the concepts and practices from \textit{Continuous Integration}, \textit{Deployment Automation}, and \textit{Monitoring \& logging} categories.  It defines the deployment pipeline stages, the automation tools,  configures this pipeline in both staging and production environments.  The expected deliveries are the documentation, implementation of the pipeline stages, continuous deployment automation, and automated tests.  Monitoring the production environment is optional.
    \item \textbf{Development Team}: implements the Product Backlog into potentially deliverable functionality by following technologies and quality criteria established. The expected deliveries are the source code of the project, unit test, communication via issues, and OSS practices execution.
    \item \textbf{Teacher}: assists in risk management, introducing theoretical concepts related to \devops, and facilitating communication with other stakeholders.
    \item \textbf{Instructor}: assists the team with training and feedback on tools, technologies, and best practices in OSS product development. They help with the culture of collaboration and share their lessons learned.
\end{itemize}

\subsection{Project Restrictions}

Project restrictions are important to limit the project's scope, set the evaluation criteria and checklist, guide planning, execution, and prioritization of tasks. It also helps instructors assess the projects' risks continually and give constant feedback to teams, according to the boundaries set by the restrictions. OSS community has a series of recommended standards focusing on building welcoming communities.  These practices aim to maintain norms, code quality, technical standards, communication, knowledge spreading, team awareness, share ownership, welcome newcomers continually. We do not cover in this paper the students' evaluation detailed evaluation criteria. Strongly leaning on OSS communities and standards, we define a set of project restrictions to guide students decision-making: 

\begin{itemize}
     \item \textbf{Project Selection}: a set of themes are available. Theme suggestion is allowed but should be approved;
    \item \textbf{Programming Language}: each team has the liberty to choose their technologies based on projects requirements;
    \item \textbf{Tests}: it is mandatory to have at least $90\%$ unit test coverage; integration tests are also mandatory. The pipeline must execute these tests automatically in stages of both continuous integration and deployment. Other tests are optional;
    \item \textbf{\devops\xspace toolset and pipeline}: students are free to choose the toolset. The following are mandatory: containerization, continuous integration, static code analysis, style sheet compliance, test automation, deployment pipeline automation, source-control branching model;
    \item \textbf{Licence}: the license must be permissive;
    \item \textbf{Documentation}:  both technical and non-technical documentation must be available on the project wiki. Obligatory documentation: vision document, architecture, deployment pipeline, product/project roadmap;
    \item \textbf{Architecture}: we recommend microservices;
    \item \textbf{Monitoring}: they should monitor the project metric. Such as test coverage, cumulative flow, knowledge evolution, risk burndown, earned value management, release train, deployments, and velocity;
    \item \textbf{Environments}: each team must have staging and production environments available;
    \item \textbf{OSS standards}: OSS community standards must be adopted. Standardization and template issues, source-control branching model, project documentation, deployment pipeline, issue communication, and feedback on Pull Requests are examples of OSS standard;
    \item \textbf{Delivery cycles}: there are two significant official releases in the course where the teams present their solutions to an evaluation board.
\end{itemize}


\section{The Course Metrics}
\label{QuantitativeAnalysis}

When correctly employing \devops\xspace concepts in a project development cycle, one expects to obtain high team productivity, a complete automated deployment pipeline, and consolidated collaboration culture. Therefore, the metrics reflecting these three aspects give a general perception of how effective a team deploys \devops\xspace concepts in practice.  

To answer the research question, we evaluate the course delivery over the years. Since 2014 \totalnumberofstudents  students enrolled in this course,  producing $47.583$ commits,  $5.255$ pull requests, and $5.820$ issues. In this section, we analyze the data from the project's repositories to verify if a project's development helps students understand and practice \devops\xspace concepts.

\subsection{Culture of Collaboration, Performance, and Productivity}

\begin{figure}[ht]
  \includegraphics[scale=0.3]{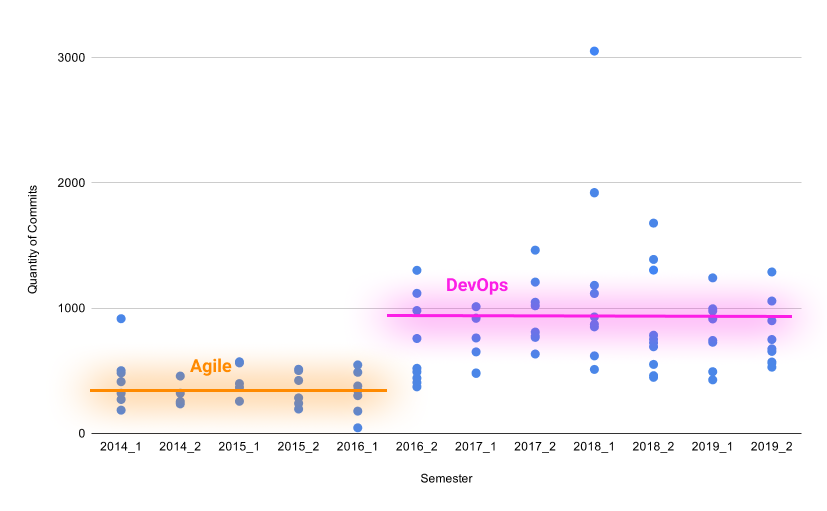}
  \caption{Commits per project from 2014 to 2019.}
  \label{fig:commits-per-project}
\end{figure}

Commit best practices are individual evaluation criteria. Architectural and project complexity are similar among projects. Therefore, the atomicity of commits is similar.  Although simplistic, we can use commits per project as a metric of team productivity, and we use it to evaluate the course setup centered? on \devops. Figure \ref{fig:commits-per-project} depicts the commits per project in this course, from 2014 to 2019. One can notice the line representing the commits overtime, highlighting the periods in which the course focused on Agile (from 2014 to 2016) and \devops\xspace (since August 2016). The improvement of teams productivity perceived overall in the course reflected adopting practices from \textit{Delivery} and \textit{Runtime} categories while updating the \textit{Project Management} category to work on full-stack teams' perspective. Even though teams could choose unfamiliar technology (programming language, frameworks) for their projects, this choice does not impact the average number of features delivered, hence productivity.

\begin{figure}[ht]
  \includegraphics[scale=0.2]{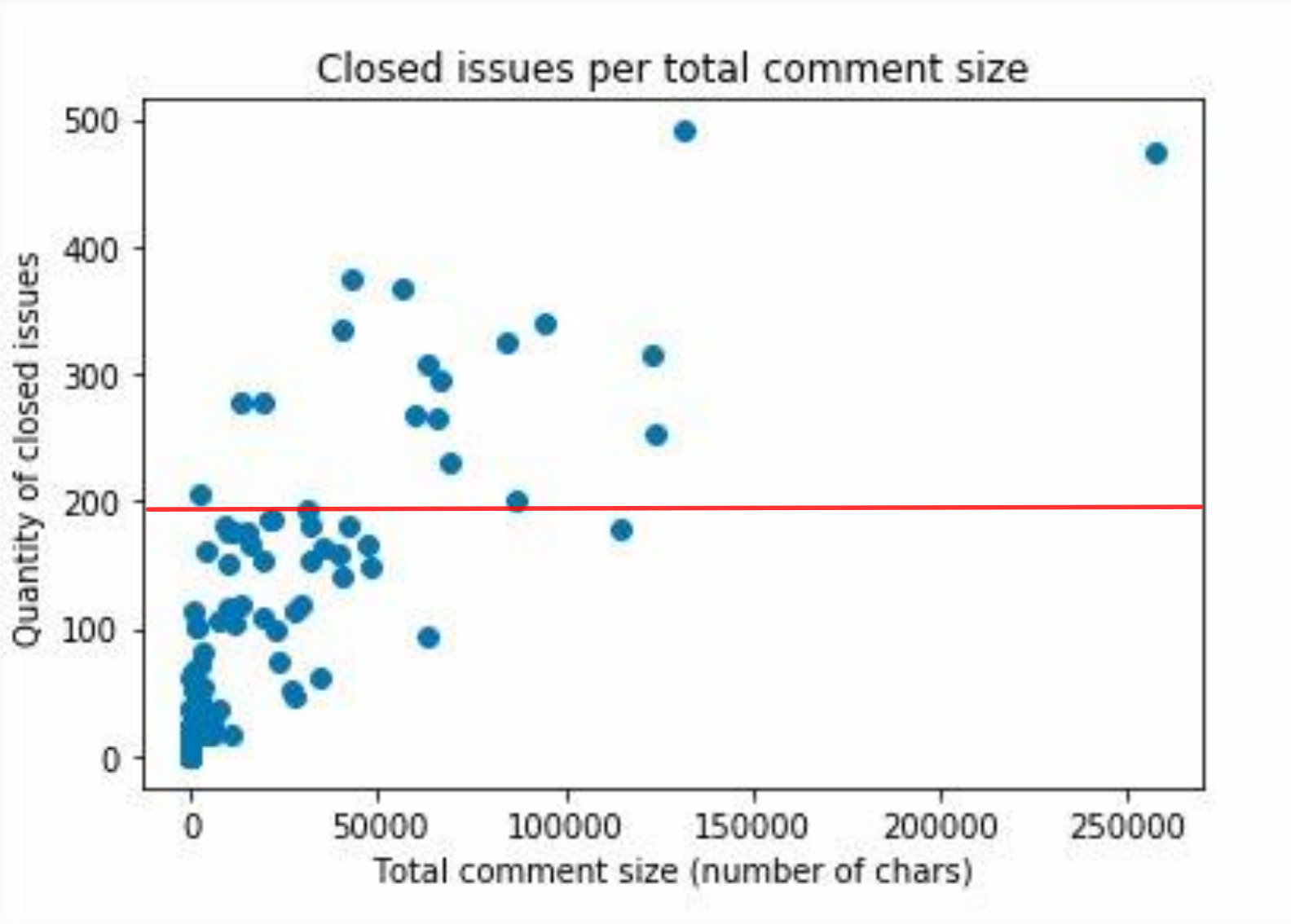}
  \caption{Culture of collaboration: impact of communication (via issue) and the teams productivity (number of closed issues)}
  \label{fig:nline-per-nissues}
\end{figure}

Efficient and frequent communication tends to impact the productivity and delivery of a team positively. A metric to evaluate the communication practice and the culture of collaboration can be the average length (number of characters) of conversations on issues. It does not reflect the communication efficiency but rather the effort to share decisions, doubts, and issues with the entire team. Another productivity or delivery metric can be the number of closed issues over time. From OSS community practice, organizations with high delivery also communicate frequently in issues. To verify this correlation in the course, Figure  \ref{fig:nline-per-nissues} shows the effect of communication practices and the teams delivery, two crucial pillars of \devops.  

In Figure \ref{fig:nline-per-nissues}, in the x-axis, we have the size of comments on issues per project, and the y-axis shows the number of issues closed per project. Each dot represents one team.  The hypothesis is that the more a team communicates in issues, they reinforce collaboration between development and operation tasks, and consequently, they have faster feature releases.
 According to their grades in class, we separate them into two groups. The teams on the upper region of the graph presented more mature continuous delivery pipelines and processes.   We compared the result in Figure \ref{fig:nline-per-nissues} with groups \textit{post mortem}, final project delivery, and grades.  Overall, we found a direct correlation with how well the group invested in collaboration, efficient communication, performance, and the final technical project delivery quality. 

\subsection{OSS, Agile, Lean Practices and Continuous Delivery}

OSS community embraced \devops\xspace practices and provide several automation tools  \cite{Leo2019}\cite{mulyana2018chatops}, and  we adopt many of them in the course, as depicted in Table \ref{tab:commands}. We continually track each team's maturity in these OSS practices. We performed a qualitative analysis on how effective the communication is in issues and pull requests, the quality and correctness of the technical communication, contribution guidelines (git branches with policies), issues and Pull Requests templates, and all technical and community documentation. Additionally, agile/lean practices such as stand-ups, planning, sprint review, pair programming, Pull request, and Kanban strengthens project and product management.

Table \ref{tab:techused} summarizes the teams' technological choices (programming languages, frameworks). There are  19 frontends and 110 backends (microservices, monolith applications, and APIs) repositories. 53\% of the projects used Docker for containerization. Containerization became obligatory in August 2018 \cite{kelly2016better}. Soon after, 95\% of repositories adopted Docker. The remaining $5\%$ projects are usually frontend repository and native apps with only documentation to configure and deploy the environment.   To configure continuous integration and deployment pipeline, groups work with shell scripting and other tools such as Rancher, Kubernetes, Travis, Jenkins, and DockerHub. As we can see, more than 50\% of the projects used shell script.

\begin{table}
\caption{Technologies used.}
\label{tab:techused}
\centering
\begin{tabular}{cccc}\hline 
\textbf{Technology} & \textbf{used on frontend} & \textbf{used on backend} & \textbf{} \\\hline 
JavaScript        & 89\%                        & 53\%                       &           \\\hline 
HTML              & 74\%                        & 47\%                       &           \\\hline 
CSS               & 68\%                        & 35\%                       &           \\\hline 
Shell             & 37\%                        & 52\%                       &           \\\hline 
Python            & 21\%                        & 55\%                       &           \\\hline 
Makefile          & 21\%                        & 25\%                       &           \\\hline 
Ruby              & 11\%                        & 15\%                       &           \\\hline 
TypeScript        & 32\%                        & 5\%                        &           \\\hline 
Objective-C       & 11\%                        & 1\%                        &           \\\hline 
Java              & 16\%                        & 7\%                        &           \\\hline 
Vue               & 16\%                        & 2\%                        &           \\\hline 
Gherkin           & 5\%                         & 6\%                        &           \\\hline 
Kotlin            & 5\%                         & 2\%                        &           \\\hline 
C\#               &                             & 2\%                        &           \\\hline 
Puppet            &                             & 1\%                        &           \\\hline 
CoffeeScript      &                             & 4\%                        &           \\\hline 
PHP               &                             & 4\%                        &           \\\hline 
LenaSYS           &                             & 1\%                        &           \\\hline 
TSQL              &                             & 1\%                        &           \\\hline 
Hack              &                             & 1\%                        &           \\\hline 
PLpgSQL           &                             & 1\%                        &           \\\hline 
Jupyter Notebook  &                             & 1\%                        &           \\\hline 
Smalltalk         &                             & 1\%                        &           \\\hline 
\end{tabular}
\end{table}

In \cite{hu2018}, Hu listed the 13 most common students' mistakes in OSS projects oriented courses. They are mostly related to coding mistakes and pull requests not following the OSS community code quality standards. When focusing on \devops\xspace, students do not usually make these mistakes because automation tools for continuous deploys, such as static code analysis, language code style, unit test automation, continuous integration, force students to employ the best programming practices. 

From the \textit{Delivery/Runtime} categories, one of the critical aspects of \devops\xspace is deployment pipeline automation. It requires knowledge of concepts like containerization, continuous integration vs. continuous deployment, pipeline stages, process (git branches and pull request policies), toolset configuration, cloud service, and microservice architecture. Typically, one student per project is responsible for studying and making decisions on deployment pipeline automation. Then, pairing with other team members to evolve and maintain the pipeline is the primary strategy to share knowledge and tools. In Table \ref{tab:Devops}, we outline how much of these concepts undergraduates in software engineers were able to implement in their projects.

\begin{table}
\caption{Deployment pipeline.}
\label{tab:Devops}
\centering
\begin{tabular}{cc}
\textbf{Pratics}       & \textbf{\% of use} \\ \hline
Containerization       & 76\%               \\
Virtualization         & 11\%               \\
None                   & 14\%               \\ \hline
Continuous Integration & 92\%               \\
None                   & 8\%                \\ \hline
Others \devops Tools   & 6\%                
\end{tabular}
\end{table}





\section{Questionnaire analysis}
\label{QualitativeAnalysis}


Before the course, only $24,1\%$ of the respondents had previous experience with OSS, and $80,3\%$ believed it was very relevant to the course. All students were under 24 years when they took the course, and $80,5\%$ were men. When asked students' expectations before attending the course, we have the following remarks:

\begin{itemize}
   \item \textbf{Developer Experience:}  students cited that the course was their first contact with the software development cycle, working in a team, learning more about coding practices and techniques. 

    \item \textbf{Work in a Real Project:}  students cited that even if they were working on capstone projects, they learned how to build software products for real problems with real potential clients. 

    
    \item \textbf{Challenge and Culture:}  It is inevitable not to notice the culture created around this course.  Students answered almost unanimously that the course is challenging, as it requires time and dedication.  One student mentioned:  \textit{"Because of the culture created around the course, I expected it to be a laborious and exhausting experience, but at the same time of great learning opportunity.}"

    \item \textbf{Project and Product Management:} They recognized the importance of management to increase their team productivity and the quality of their artifacts.
    
    \item \textbf{Development, Delivery, and Runtime:} The students stated they mature on their understanding of the problems associated with software delivery, the importance of automation,  versioning control, and clear acceptance criteria to new features.
\end{itemize}



With a focus on \devops, students' most relevant technical aspects are the design and implementation of a continuous deployment pipeline. We ask them to rank the most relevant tools and evaluate several \devops\xspace concepts/practices while executing their project. Continuous Integration ($87\%$) is the most critical in the students' opinion, followed by Continuous Deploy ($79.6\%$), Code Versioning ($80.6\%$), Continuous Delivery ($70.4\%$), Environment Isolation ($69.4\%$), and Build Automation with $63\%$. 

These concepts introduced from the beginning of the course, added with tips and tools already used in previous classes, increase discussions and understanding become more mature and straightforward to apply. Automation, Product Vision, Feedback, Continuous Delivery, and Containerization were the \devops\xspace practices that added most value from students' perspective as "Very relevant". Aspects like Packaging ($27.8\%$) and Architecture Microservices ($30.6\%$) are at the bottom of the ranking, as these aspects were recently added to the course. 

About the challenges, $27\%$ reported the difficulty in both team and self-time managing. They report they have little time to understand  and implement a large number of concepts, tools, techniques. They also recognized the difficulties in guaranteeing an efficient communication. Numerous students stated the following challenge: "\textit{Manage and remove impediments from the development team to not impact delivery. And surely keep communication between everyone in the group clear.}" (Student)

\begin{figure}[ht]
  \includegraphics[scale=0.50]{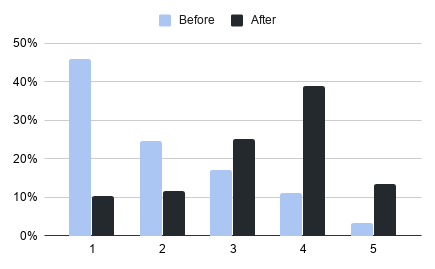}
  \caption{Average of the technologies and concepts used, perception of knowledge before and after the course.}
  \label{fig:average-confidence-before-after}
\end{figure}

We asked students their perception of their prior and post-course knowledge in  coding techniques, programming language and \devops\xspace practices. They did a self-assessment on a scale of 1 to 5. Figure \ref{fig:average-confidence-before-after} presents an average of the responses of the leading technologies and concepts used (Python, JavaScript, Web Programming, and Agile Practices) before and after the course. Before the course, the majority of responses were 1, showing that they did not have a previous contact to these programming languages and practices.After finishing the course their  self-assessment improved in general to  3 and 4.  As it is a self-assessment, we cannot say that students' knowledge increased, but we can infer that students feel more confident about the tools, concepts, and technologies after working with them during the course.





\section{Lessons Learned}
\label{SecFurtherAnalysis}


In this section, we discuss some of the lessons learned observed from data analysis, insights from teachers, and feedback from people and partners who have already gone through the discipline. As it is an experimental discipline, the culture of continuous feedback was adopted, collecting through surveys and meetings with students at the end of each semester, thus applying improvements to the next offer. This section aims primarily to present some of the issues professors, instructors, and students supporting \devops\xspace teaching will likely face.

\begin{itemize}

\item \textbf{Culture is a key factor} -  it is reported the importance of \devops\xspace culture to make continuous deployment effective. The same is true for teaching it. It is essential to develop and foster a safe environment for the students to learn and fail,  reward open collaboration, open communication, engaged mentoring, and technical leadership, self-organization, team-oriented over individuals.
With the commitment to keep technical communication centralized on issues, the teams organized themselves to maintain this culture using bots on communication channels and other tools to maintain decentralized communication and the entire community. Figures \ref{fig:nline-per-nissues}, \ref{fig:commits-per-project} and Table \ref{tab:techused} indicate that performance of every team is quite similar and it is not much affected by the technical choices, a reflection the culture of collaboration. Students are aware of the importance of this particular course in their professional qualification, and it motivates them to engage and dedicate themselves throughout the entire project.  

\item \textbf{Encourage the use of different technologies} - students freedom to choose their technological setting, alongside culture, instigate them to experiment with a new toolset, new languages, new frameworks. Table \ref{tab:techused} show all the language and frequency used by the teams. The adopted technologies are continuously evolving, and the students choice are compatible with trending technologies adopted by practitioners in Stackoverflow \footnote{https://stackoverflow.com}. In this process,  professors emphasize that concepts are more important than tools, and each technical choice should be attached to a \devops\xspace practice and the problem being solved.  Each team must justify their technological choice based on such concepts and how they enable \devops. As stakeholders, professors and instructors share concerns, past project mistakes, known risks, and, finally, approve team choices. This freedom enables innovation and keeps the course updated with current practices and toolsets used by the \devops\xspace community.

\item \textbf{No detailed plan for achieving the final delivery} -  the course does not provide a roadmap to students to follow through the project, which is similar to real software projects. Leaving this core decision to the teams exercise their management and decision-making skills, but it also encourages them to collaborate with other teams, with their mentor, with the professor. The project roadmap encompasses the complexity of a software project when focusing on continuous delivery and \devops: it involves defining the process, the toolset, the delivery pipeline, the automation, the runtime environment, monitoring and assigning people attributions, the deliverables, the team training, among others. All these choices combined impact the effectiveness of \devops\xspace in a controlled course environment. It gives the students a systematic view of \devops\xspace aspects from experience. Learning \devops\xspace from experience is more important and beneficial than a purely theoretical approach because it evidences how both technical and non-technical \devops\xspace concepts are interconnected, and they both impact the final delivery. Several students reported their initial struggle to make decisions, their insecurity to not have a pre-defined list of tasks.  Nevertheless, they recognized the importance of decision-making to comprehend the elements that impact continuous delivery.

\item \textbf{OSS communities practices} - There is still significant debate among practitioners and managers about \devops; the community standards, tools, and practices guidelines are continually evolving. There is a variety of suggested pipeline stages, depending on the application context.  \devops\xspace tools and practices are rapidly changing, and most of them are open source, and the use of such tools directly impacts developers and operators' responsibilities. Table \ref{tab:Devops} also highlights the use of other practices more directly linked to integration and continuous delivery in projects. Thus, it is fundamental to project-based \devops\xspace courses to request teams to employ the current OSS  and \devops\xspace community standards (documentation, communication, process, tools) to keep the discussions and concepts up to date. 

\item \textbf{Students portfolio and collaboration with local software industry} - it was well documented the benefits for project-oriented OSS courses to build student portfolio. Like Outreachy, Google Summer of Code, many industry initiatives, finance undergraduates in contributing to open source communities. They are both strategies to bridge the gap between academy and industry. In countries where the software industry is not well established or does not have enough investments to follow global trends, we hypothesize that the initiative to teach undergraduates current \devops\xspace concepts in a practical setup promotes innovation to the local industry.  These students will build globally competitive skills, and they can share them with the local industry.

\item \textbf{Mentoring} - we have tested several mentoring formats: two instructors per team participating continually and directly in project decisions and responsible for team coaching; both technical and non-technical collaborators from local industry working as mentors. Each format has its benefits and disadvantages. External collaborators can have low commitment to guide students; instructors chosen by the professor may not have the necessary technical skill necessary to mentor a particular project; non-proactive teams may not search for mentors. These risks are always present, and projects can be adversely affected by them. However, the benefits of any format of mentoring are countless \cite{Trainer:2017}. Mentors aid the professor in assessing project risks, and they bring different perspectives, practices, and technologies that help bridge the gap between academia and the software industry.

\item \textbf{Team versatility is important} -  It is impractical for every student in a team to develop all of the necessary skills. \devops\xspace requires versatile and skilled developers, managers, and product strategists. Therefore, professors and instructors should encourage students to assume the responsibilities and tasks that correspond to their interests. It is essential to have students with the ability to understand many aspects of the system (generalists). Additionally, another skill necessary in learning \devops\xspace is the ability to be responsible (and accountable) for making deployment decisions, taking quality, performance, acceptance criteria into account. Dealing with unpredicted scenarios and risks and taking appropriate initiatives when necessary. Even when students take managerial tasks, they should be aware that technical management is essential, and they should aim to influence the team rather than give orders.

\item \textbf{Evaluation should not focus only on deliverable} - even though it is possible to monitor metrics on \devops, it is vital for professors not to focus strictly on project/product metrics. Risk management maturity,  process experience, communication effectiveness,  team engagement, innovation should also be considered in students' evaluation. Otherwise, it will not encourage students to experiment and fail,  try new frameworks and tools,  and take risks. These qualitative criteria must be explicit at the beginning of the course to build trust among the stakeholders (students, professors, instructors).  Even though we did not cover the course evaluation in the current paper, some of the practices depicted in Table \ref{tab:commands} are not measurable. 

\item \textbf{External Evaluators} - It is beneficial to invite externals to evaluate and give teams feedback from both industry and academy perspective. They have unique views and might bring new questioning to the course, suggesting improvements,  exposing blind spots and bias not perceived by professors and instructors supervising the course. Since \devops\xspace is inherently multidisciplinary, so should be these externals evaluators. Designers, SRE experts, software architects, open source advocates, developers are examples of past evaluators. They are essential to enable the course to improve continually. External evaluators continually review the practices covered in Table \ref{tab:commands}.

\end{itemize}

\section{Related works}
\label{SecRelatedWorks}

Because of the novelty of \devops, few papers on \devops\xspace education were published, and it is still an open challenge \cite{Leo2019}, and most of the papers treat teaching \devops\xspace in Master Degrees and specializations. Chung et al. \cite{Chung:2016} present the knowledge, skills, and abilities to \devops\xspace education, such as Architectural pattern, automation, unit testing, productivity, and agile development, and how they are relevant in web development courses. Christensen  \cite{Christensen:2016} claims teaching \devops\xspace is challenging due to its skill-focused competence nature, that goes through orient-object, testing, database subjects, and is a response for problems like speed, scale, and availability. New curriculum based on \devops\xspace concepts and practices are proposed in \cite{Capozucca2019}\cite{devopscv:segey2020}. Both papers outline the importance of project-oriented courses in the Software Engineering curriculum. Capozucca \emph{et al.} \cite{Capozucca2019} argue that in this pedagogical methodology, every team in a course should be assigned to the same project in order to enable consistency when evaluating groups' work. We disagree with this conclusion once our results have indicated that different projects motivate teams and foster innovation. The major challenge we faced in our context is to define quantitative metrics to evaluate non-technical aspects of the project development. Diehl in \cite{Baltes_2018} characterizes the skills necessary to master software development, and they evidence the relevance of mentoring to build knowledge and thus contribute to the improvement of expertise. Pang et al.  \cite{pang2020} used Grounded Theory (GT) to study DevOps education from both academic and industrial perspectives. They concluded that institutions were not teaching students about \devops, because academics were not interested in adopting or teaching \devops. In this paper, we employ most of the \devops \xspace Education Hypotheses proposed in \cite{pang2020}.

Bringing an OSS project into a teaching context is not uncommon. According to \cite{carrigton2003}, it helps students understand the differences between their small projects and real-life large scale software.  Pinto et al. \cite{Pinto:2019}\cite{Pinto:2017} present both teacher and students' perspectives on using OSS projects to learn more advanced software engineering practices and techniques while maintaining and evolving OSS. Students felt motivated to participate in a real project, point out that the subject was a starting point in the OSS community and an opportunity to build a portfolio. Besides, some students became active OSS contributors. 
\section{Limitations}

This study has limitations, and we strove to mitigate their potential impacts. First, our study is limited by one university and the number of students who participated in our study. This empirical research took \coursetime years, with four of them were \devops\xspace oriented. We use coding to identify OSS, agile and lean practices to assign to \devops\xspace concepts. Some may argue that coding is a personal interpretation, and the validity of coding is questionable. To address this concern, both authors performed the coding and reviewed it by external invited evaluators at every final course cycle.

\section{Concluding Remarks}
\label{SecConclusion}


\devops\xspace education is more challenging than any other kind of technical training. \devops\xspace is much more than a specific technology, it is a set of technical and non-technical concepts. Adopting a given practice or implementing a pipeline stage automation has a nonlinear effect on continuous delivery (it is a complex system).  In this paper, we present a project-oriented \devops\xspace course, and we introduce DevOps concepts through Agile, Lean, and Open source practices and tools. From the \devops\xspace conceptual map presented in \cite{Leo2019}, we selected the concepts related to the software development cycle. Then, we apply the coding technique to assign Agile, Lean, and OSS practice to these concepts, depicted in Table \ref{tab:commands}. In a project-oriented course, students employ these practices and contribute to a capstone OSS project. 

Automating the deployment pipeline does not necessarily guarantee continuous delivery\cite{humble2010continuous}. Students should experience this complexity, perceiving that continuous delivery is achieved by combining practices and automation while optimizing resources and identifying waste. It provides undergraduates an experience much closer to a real software project's challenges and gives a more realistic perception of these concepts' importance.

To evaluate this course scenario, we mined \numberofrepositories repositories and evaluated this course over \coursetime years.  We also surveyed to assess students' perceptions of \devops\xspace concepts and practices. From the repository of the project, we wanted to understand if the OSS, agile and lean practices/tools brought the benefits of  \devops\xspace adoption. We analyzed the teams' productivity, the culture of collaboration, and the automation pipeline. We observed that the adoption of \devops\xspace over a previous agile approach in class improved overall team productivity. From the survey, we observed students tend to assign more relevance to technical \devops\xspace practices, such as continuous integration and deployment, over non-technical practices, like communication and documentation. We believe it is because they value practices that provide instant or measurable benefits.



\section*{Acknowledgements}  

We want to thank all the teachers, instructors, external evaluators/stakeholders, and students who support the course development and improve it continually. The complete list of students, contributors can be found in the course organization \url{\organizacagit}.

\bibliographystyle{ieeetr}
\bibliography{bibliography}

\end{document}